\documentclass[12pt]{amsart}

\usepackage[foot]{amsaddr}
\usepackage{graphicx}
\usepackage{amssymb}
\usepackage{amsmath}
\usepackage{amsthm}
\usepackage{a4wide}
\usepackage{url}
\usepackage{color}
\usepackage{algorithm}
\usepackage{algpseudocode}
\usepackage{hyperref}
\usepackage{enumerate}
\usepackage{csquotes}

\theoremstyle{plain}
\newtheorem{theorem}{Theorem}[section]
\newtheorem{lemma}[theorem]{Lemma}
\newtheorem{corollary}[theorem]{Corollary}

\theoremstyle{definition}

\newtheorem{remark}[theorem]{Remark}

\newcommand{\pf}{\noindent{\em Proof:}}
\newcommand{\epf}{\hfill\hbox{\rule{3pt}{6pt}}\\}

\title{Labeling and folding multi-labeled trees}

\author{Vincent Moulton}
\address[V. Moulton]{School of Computing Sciences, University of East Anglia, UK}
\email{v.moulton@uea.ac.uk}
\author{Andreas Spillner}
\address[A. Spillner]{Merseburg University of Applied Sciences, Germany}
\email{andreas.spillner@hs-merseburg.de}


\date{\today}

\begin{document}

\begin{abstract}
In 1989 Erd{\H{o}}s and Sz{\'e}kely showed that
there is a bijection between (i) the set of rooted
trees with \(n+1\) vertices whose leaves are bijectively labeled
with the elements of \([\ell]=\{1,2,\dots,\ell\}\) for some \(\ell \leq n\),
and (ii) the set of partitions of \([n]=\{1,2,\dots,n\}\).
They established this via a labeling algorithm
based on the anti-lexicographic ordering of 
non-empty subsets of \([n]\)
which extends the labeling of the leaves of a given tree
to a labeling of all of the vertices of that tree. 
In this paper, we generalize their approach by 
developing a labeling algorithm for {\em multi-labeled trees}, that is,
rooted trees whose leaves are
labeled by positive integers but in which 
distinct leaves may have the same label. 
In particular, we show that certain orderings
of the set of all finite, non-empty multisets
of positive integers can be used to 
characterize partitions of a multiset 
that arise from labelings of multi-labeled trees.
As an application, we show that
the recently introduced class of labelable phylogenetic networks
is precisely the class of phylogenetic networks that are stable
relative to the so-called folding process on multi-labeled trees. We also 
give a bijection between the labelable phylogenetic networks with leaf-set $[n]$
and certain partitions of multisets. 
\end{abstract}

\maketitle

\section{Introduction}
\label{sec:intro}

\begin{figure}
\centering
\includegraphics[scale=1.0]{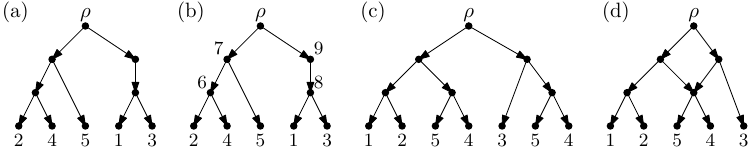}
\caption{(a) A semi-labeled tree with~\(\ell=5\) and~\(n=9\).
(b) The semi-labeled tree in (a) with vertices 
bijectively labeled with the elements of \([9]\).
Taking for each non-leaf vertex~\(v\) the subset of
\([9]\) labeling the children of~\(v\), yields the
partition \(\{\{1,3\},\{2,4\},\{5,6\},\{8\},\{7,9\}\}\) of~\([9]\).
(c) A multi-labeled tree.
(d) The phylogenetic network obtained by folding the multi-labeled
tree in~(c).}
\label{fig:intro}
\end{figure}

A \emph{semi-labeled tree} is a rooted tree in 
which the leaves are bijectively labeled with the
elements of~\([\ell]=\{1,2,\dots,\ell\}\), for
$\ell$ some positive integer 
(see e.g. Figure~\ref{fig:intro}(a)).
In the ground-breaking work~\cite{erdHos1989applications},
Erd{\H{o}}s and Sz{\'e}kely
established a bijection between the set of partitions
of~\([n]=\{1,2,\dots,n\}\) and the set of 
semi-labeled trees with \(n+1\) vertices.
Their proof relies on 
(i)~giving an algorithm for labeling the interior vertices of
a semi-labeled tree that extends the leaf labeling and
(ii)~using the simple observation that, given a rooted
tree $T$ with~\(n+1\) vertices
in which all vertices other than the root
are bijectively labeled with the elements of \([n]\),
a partition of~\([n]\) is obtained by taking,
for each non-leaf vertex~\(v\) of~\(T\), the subset of~\([n]\)
labeling the children of~\(v\)
(see e.g. Figure~\ref{fig:intro}(b)). A similar 
approach yields the bijection between matchings and
binary rooted \emph{phylogenetic trees} 
(semi-labeled trees in which 
every interior vertex has two children)
described in~\cite{diaconis1998matchings}, and is also used to devise 
phylogenetic tree encodings \cite{chauve2025vector,penn2025phylo2vec,richman2025vector}. 
These types of results have subsequently been used 
to, for example, navigate tree space \cite{penn2025phylo2vec},
and place a semi-group structure on the set of binary rooted 
phylogenetic trees~\cite{francis2022brauer}.

In this paper,
we investigate to what extent the 
results in \cite{erdHos1989applications} can be extended to rooted trees
(and networks) whose leaves are labeled by positive integers
but in which distinct leaves may have the same label
(see e.g. Figure~\ref{fig:intro}(c)). 
Such trees arise in phylogenetics under the name of \emph{multi-labeled trees}
(see e.g. \cite{huber2006phylogenetic}).
We shall show that such trees are in bijection with certain partitions of
multisets. As with  Erd{\H{o}}s and Sz{\'e}kely's result, 
the key to establishing this bijection is to develop an algorithm that
extends the labeling of the leaves of a multi-labeled tree
to a labeling of all vertices of the tree
in such a way that the tree and its labeling can be uniquely 
recovered from the resulting multiset partition. Our labeling
algorithm is parameterized by {\em any} given ordering~\(\preceq\) of the set of all
finite, non-empty multisets of positive integers. 
Moreover, choosing~\(\preceq\) to be the \emph{anti-lexicographic ordering}, 
produces the same output as the algorithm used in~\cite{erdHos1989applications} 
when applied to a semi-labeled tree.

We shall also describe an application of our multi-labeled tree
results to {\em phylogenetic networks}, graphs that are used in
phylogenetics to represent complex evolutionary
relationships (see e.g. \cite{S16a}). 
For our purposes, phylogenetic networks
are essentially directed acyclic graphs whose leaves are bijectively labeled 
by the elements of $[\ell]$ for some $\ell \ge 1$  
(cf. Section~\ref{sec:network:classes} for a full definition).
Recently the class of \emph{labelable} phylogenetic networks was introduced
in~\cite{francis2023labellable}. These networks
are defined by the property that a certain variant of the labeling
algorithm from~\cite{erdHos1989applications} extends the labeling
of the leaves in the network to a labeling of all vertices
so that distinct vertices have distinct labels.
A family of label subsets can then still be 
obtained from the children of internal
vertices of the network as we did for the tree in
Figure~\ref{fig:intro}(b), but the subsets need no longer be disjoint,
and form a so-called \emph{expanding cover} rather than a partition.

Based on our labeling algorithm, we establish 
an alternative characterization of labelable phylogenetic networks based on
a process used to create phylogenetic networks  
called \emph{folding} \cite{huber2006phylogenetic}.
This process identifies and overlays isomorphic subtrees
in a multi-labeled tree in a systematic way
so as to produce a phylogenetic network
in which distinct leaves have
distinct labels (see e.g. Figure~\ref{fig:intro}(d)).
The reverse process, called \emph{unfolding},
turns a phylogenetic network into a multi-labeled tree.
Note that unfolding a phylogenetic network
and then folding the resulting multi-labeled tree may yield
a phylogenetic network that is not isomorphic to the original network; in case
it is isomorphic the network is called {\em stable} \cite{huber2016folding}.
Here we prove that the labelable phylogenetic networks
are precisely the stable networks.

We also consider the problem of characterizing when a partition of a multiset
corresponds to a multi-labeled tree relative to some 
ordering \(\preceq\)  of the set of all
finite, non-empty multisets of positive integers. 
Interestingly, this appears to be a difficult problem 
for an arbitrary ordering~\(\preceq\). 
In~\cite{erdHos1989applications} 
this issue is overcome by selecting some specific ordering.
Here we show that it is possible to characterize 
when a partition of a multiset
corresponds to a multi-labeled tree for 
a more general class of orderings that contains the 
anti-lexicographic ordering.
In this way, we are then able to characterize
when a partition of a multiset corresponds to 
certain subclasses of stable phylogenetic networks. 
Note that in~\cite{francis2024phylogenetic}
various classes of phylogenetic networks are 
characterized by properties of their corresponding expanding covers.
Our approach therefore provides a complementary way to characterize certain phylogenetic network classes.

The rest of this paper is structured as follows.
In Section~\ref{sec:full:labelings} we first provide formal definitions
of some of the concepts informally described in this
introduction and then present our labeling algorithm
(Algorithm~\ref{alg:label:network}).
Then, in Section~\ref{sec:partitions:from:trees}, we
show how multi-labeled trees are uniquely determined by their
associated partitions of multisets (Theorem~\ref{thm:alg:reconstruct:tree}). 
In Section~\ref{sec:char:tree:generated:partitions} we first discuss
the role of the ordering~\(\preceq\) used in Algorithm~\ref{alg:label:network}
leading to a description (Lemma~\ref{lem:char:labeling:consistent})
of those orderings
that allow to characterize the partitions of multisets that
correspond to multi-labeled trees (Theorem~\ref{char:exp:partitions}).
In Section~\ref{sec:folding:unfolding} we establish the
equivalence of labelable and stable networks
(Theorem~\ref{thm:stable:networks}).
Based on this, in Section~\ref{sec:network:classes},
we illustrate how some classes of phylogenetic networks can be
characterized by properties of partitions of multisets. We 
conclude with some remarks on potential 
future  directions and applications of our results.

\section{Labelings and the labeling algorithm}
\label{sec:full:labelings}

In this section we present our labeling algorithm. We 
begin with some definitions.
A \emph{rooted network} \(\mathcal{N}=(G,\rho)\) is a directed
acyclic graph (DAG)~\(G\) with a unique vertex~\(\rho\) of
in-degree~0. \(\rho\)~is called the \emph{root} of~\(\mathcal{N}\).
A \emph{leaf} of~\(\mathcal{N}\)
is a vertex of out-degree~0 in~\(G\). We denote the set
of leaves of~\(\mathcal{N}\) by~\(L(\mathcal{N})\).
A vertex in~\(G\) that is not a leaf is an \emph{interior vertex}
of~\(\mathcal{N}\). Let~\(u\) be a vertex in a rooted
network~\(\mathcal{N}=((V,E),\rho)\). Then
the set of all vertices~\(v\) with \((u,v) \in E\) is denoted
by~\(C_u\) and the vertices in~\(C_u\) are called the
\emph{children} of~\(u\). Note that a vertex in a rooted network
is allowed to have in-degree and out-degree both equal to~1.

\begin{figure}
\centering
\includegraphics[scale=1.0]{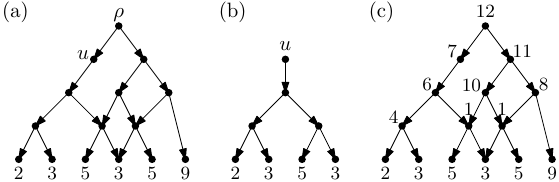}
\caption{(a) A leaf-labeled network~\((\mathcal{N},\lambda)\).
(b) The leaf-labeled network~\((\mathcal{N}_u,\lambda_u)\) for the
vertex~\(u\) in (a). 
(c) The fully-labeled network
produced by Algorithm~\ref{alg:label:network} from the leaf-labeled
network in (a) using as~\(\preceq\) the lexicographic ordering.}
\label{fig:ex:leaf:labeled:network}
\end{figure}

Let \(\mathbb{N}^* = \mathbb{N} - \{0\}\) and, for any
\(n \in \mathbb{N}^*\), let \([n] = \{1,2,\dots,n\}\).
A \emph{leaf labeling} of a rooted network \(\mathcal{N}\)
is a map \(\lambda: L(\mathcal{N}) \rightarrow \mathbb{N}^*\).
Note that we don't require
that~\(\lambda\) is injective. A \emph{leaf-labeled network}
\((\mathcal{N},\lambda)\) consists of a rooted network~\(\mathcal{N}\)
and a leaf labeling~\(\lambda\) of~\(\mathcal{N}\)
(see Figure~\ref{fig:ex:leaf:labeled:network}(a) for an example).
Two leaf-labeled networks \(((G=(V,E),\rho),\lambda)\) and
\(((G'=(V',E'),\rho'),\lambda')\) are \emph{isomorphic} if there
exists a DAG-isomorphism \(f: V \rightarrow V'\) such that
\(\lambda(v) = \lambda'(f(v))\) for all \(v \in L(G,\rho)\).
A leaf-labeled network~\((\mathcal{N},\lambda)\)
is called a \emph{leaf-labeled tree} if all vertices in
\(\mathcal{N}\) have in-degree at most~1.

We remark that the definition of leaf-labeled networks
may appear slightly more general
than necessary. The motivation for this is twofold: We want it to
include all the types of networks mentioned in the introduction
(i.e. semi-labeled trees, multi-labeled trees and phylogenetic
networks) and we want to avoid certain technical problems when
considering induced subnetworks. To formally define the latter,  
let~\(u\) be a vertex in a leaf-labeled
network \((\mathcal{N}=(G=(V,E),\rho),\lambda)\).
Then the leaf-labeled network \((\mathcal{N}_u,\lambda_u)\) 
consists of (see Figure~\ref{fig:ex:leaf:labeled:network}(b) for an example)
\begin{itemize}
\item
the rooted network~\(\mathcal{N}_u=(G_u,u)\) with root~\(u\)
that is obtained from the sub-DAG~\(G_u\) of~\(G\) induced by the set
of those vertices \(v \in V\) for which there exists a
directed path from~\(u\) to~\(v\) in~\(G\), and
\item
the restriction~\(\lambda_u\) of the map~\(\lambda\) to the
set of leaves of~\(\mathcal{N}_u\).
\end{itemize}

Given a leaf-labeled network~\((\mathcal{N}=(G=(V,E),\rho),\lambda)\),
we want to extend the leaf labeling~\(\lambda\) to a
map \(\phi: V \rightarrow \mathbb{N}^*\).
Such a map~\(\phi\) is called a \emph{full labeling} of~\(\mathcal{N}\) and
\((\mathcal{N},\phi)\) is referred to as a 
\emph{fully labeled network}. 
In addition, for an interior vertex~\(u\) of a fully labeled network
\((\mathcal{N},\phi)\), let \(F_{(u,\phi)} = \{\phi(v) : v \in C_u\}\) 
denote the multiset of labels assigned by \(\phi\) to
the children of~\(u\). 

A particular full labeling is obtained by Algorithm~\ref{alg:label:network}.
Let \(\mathbb{M}\)
denote the set of all finite, non-empty multisets whose elements
are from the set~\(\mathbb{N}^*\). 
For a multiset \(M \in \mathbb{M}\) and \(x \in \mathbb{N}^*\),
the multiplicity of~\(x\) in~\(M\) is denoted by~\(M(x)\),
with \(M(x) = 0\) indicating that~\(x\) is not contained in~\(M\).
Algorithm~\ref{alg:label:network}
is provided with an ordering~\(\preceq\) of the multisets
in~\(\mathbb{M}\) (we write \(M \prec M'\) for \(M,M' \in \mathbb{M}\)
if \(M \preceq M'\) and \(M \neq M'\)).
The full labeling produced by Algorithm~\ref{alg:label:network} 
for a given leaf-labeled network \((\mathcal{N},\lambda)\) 
depends on the choice of~\(\preceq\) and is denoted
by~\(\phi_{(\lambda,\preceq)}\).
In the literature, various orderings of the multisets
in~\(\mathbb{M}\) have been considered
(see e.g. \cite{martin1989geometrical}).
This includes the \emph{lexicographic ordering} defined by putting
\(M \preceq M'\) for \(M,M' \in \mathbb{M}\)
if either \(M=M'\) or \(M(x) < M'(x)\)
for the smallest \(x \in \mathbb{N}^*\) with \(M(x) \neq M'(x)\).
An example of the output of Algorithm~\ref{alg:label:network}
is shown in Figure~\ref{fig:ex:leaf:labeled:network}(c).

\begin{algorithm}
\caption{Labeling the interior vertices of a leaf-labeled
network~\((\mathcal{N},\lambda)\)}
\label{alg:label:network}
\begin{algorithmic}[1]
\Procedure{ComputeFullLabeling}{$\mathcal{N}$,$\lambda$}
\State $\phi(v) \gets \lambda(v)$ for all $v \in L(\mathcal{N})$ \Comment{labels of leaves are given}
\State $\mathbb{L} \gets \lambda(L(\mathcal{N}))$ \Comment{set of labels currently used}
\State $U \gets$ set of interior vertices of~$\mathcal{N}$ \Comment{set of currently unlabeled vertices}
\While{$U \neq \emptyset$}
   \State $W \gets \{u \in U \,:\, \text{all children of} \ u \ \text{have been labeled by} \ \phi\}$
   \State $W' \gets \{w \in W \,:\, F_{(w,\phi)} \ \text{is minimum with respect to} \ \preceq\}$
   \State $k \gets \min ( \mathbb{N}^* - \mathbb{L} )$ \Comment{select value used as next label}
   \State $\phi(w) \gets k$ for all $w \in W'$ \Comment{label all vertices in $W'$}
   \State $\mathbb{L} \gets \mathbb{L} \cup \{k\}$ \Comment{update set of labels currently used}
   \State $U \gets U - W'$ \Comment{update set of unlabeled interior vertices} 
\EndWhile
\EndProcedure
\end{algorithmic}
\end{algorithm}

Note that Algorithm~\ref{alg:label:network} is related to
a (reverse) \emph{topological sort} of the vertices of a DAG
(see e.g. \cite[Sec.~20.4]{cormen2022introduction}).
A DAG has, in general, many different orderings of its
vertices that form a valid output of a topological sort.
The ordering~\(\preceq\) on~\(\mathbb{M}\) used in
Algorithm~\ref{alg:label:network} essentially guides
which of these orderings is chosen. 
In the following, we state some key properties of
Algorithm~\ref{alg:label:network}.

\begin{figure}
\centering
\includegraphics[scale=1.0]{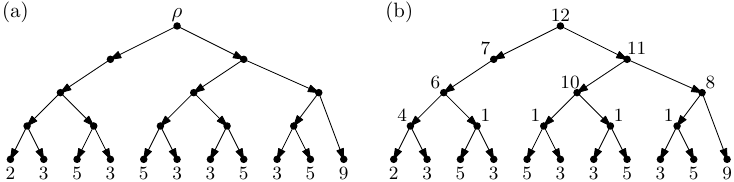}
\caption{(a) A leaf-labeled tree \((\mathcal{T},\lambda)\).
(b) The full labeling \(\phi=\phi_{(\lambda,\preceq)}\)
of the leaf-labeled tree in (a)
produced by Algorithm~\ref{alg:label:network}
using as~\(\preceq\) the lexicographic ordering. Interior vertices \(u,v\)
with the same label have isomorphic leaf-labeled
trees \((\mathcal{T}_u,\lambda_u)\) and \((\mathcal{T}_v,\lambda_v)\)
(cf. Lemma~\ref{lem:label:tree:isomorphic}).}
\label{fig:iso:same:label}
\end{figure}

\begin{lemma}
\label{lem:label:tree:isomorphic}
Let \((\mathcal{T}=((V,E),\rho),\lambda)\) be a leaf-labeled tree and
\(\phi = \phi_{(\lambda,\preceq)}\) be the full labeling of \(\mathcal{T}\)
produced by Algorithm~\ref{alg:label:network}.
Then, for any two interior vertices \(u,v \in V\), the leaf-labeled
trees \((\mathcal{T}_u,\lambda_u)\) and \((\mathcal{T}_v,\lambda_v)\)
are isomorphic if and only if \(\phi(u) = \phi(v)\)
(see Figure~\ref{fig:iso:same:label} for an example).
\end{lemma}

\pf
Consider interior vertices \(u,v \in V\).
Since Algorithm~\ref{alg:label:network} is deterministic,
\((\mathcal{T}_u,\lambda_u)\) being isomorphic to
\((\mathcal{T}_v,\lambda_v)\) immediately implies \(\phi(u) = \phi(v)\).

Conversely, assume for a contradiction
that there exist interior vertices
\(u,v \in V\) such that \(\phi(u) = \phi(v)\)
but \((\mathcal{T}_u,\lambda_u)\) is not isomorphic
to \((\mathcal{T}_v,\lambda_v)\). 
Consider the first iteration of the while-loop of
Algorithm~\ref{alg:label:network} where such two
vertices are assigned their labels.
Since \(\phi(u) = \phi(v)\) the multiset of
labels of the children of \(u\) and \(v\) must
coincide. But this implies, by the choice of
the iteration, that there is a one-to-one
correspondence between the leaf-labeled
trees \((\mathcal{T}_w,\lambda_w)\), \(w \in C_u\),
and the leaf-labeled
trees \((\mathcal{T}_{w'},\lambda_{w'})\), \(w' \in C_v\),
such that the corresponding leaf-labeled
trees are isomorphic. It follows that
\((\mathcal{T}_u,\lambda_u)\) is isomorphic
to \((\mathcal{T}_v,\lambda_v)\), a contradiction.
\epf

The following can be regarded as being an analogue
of~\cite[Thm.~3.3]{francis2023labellable}
and will form the basis for the
definition of labelable networks
(see Section~\ref{sec:folding:unfolding}).

\begin{theorem}
\label{thm:labeling:injective}
Let \((\mathcal{N}=((V,E),\rho),\lambda)\) be a leaf-labeled network.
Then the following are equivalent:
\begin{itemize}
\item[(a)]
The full labeling \(\phi = \phi_{(\lambda,\preceq)}\)
produced by Algorithm~\ref{alg:label:network} is injective.
\item[(b)]
The leaf labeling \(\lambda\) is injective and
\(C_u=C_v\) implies \(u=v\) for all \(u,v \in V\).
\end{itemize}
\end{theorem}

\pf
Clearly, (a) implies (b). So assume that (b) holds.
Assume for a contradiction that there exist 
two distinct vertices \(u,v \in V\) with
\(\phi(u)=\phi(v)\). Consider the first iteration
of the while-loop of
Algorithm~\ref{alg:label:network} where such two
vertices are assigned their labels.
By the choice of the iteration,
the children of \(u\) and \(v\), respectively,
have pairwise distinct labels. Thus, since
\(C_u \neq C_v\), the set of labels of the children
of \(u\) and \(v\) are different. But then
\(u\) and \(v\) are not labeled in the same
iteration, a contradiction. 
\epf

\section{Partitions from leaf-labeled trees}
\label{sec:partitions:from:trees}

In the last section we presented Algorithm~\ref{alg:label:network}
which, in particular, can be applied to any leaf-labeled
tree~\((\mathcal{T},\lambda)\) to produce a full labeling
of that tree. In this section, we first describe how a
partition of a multiset can be obtained from this full labeling and
then present an algorithm (Algorithm~\ref{alg:mulfromtreepartition})
that reconstructs the leaf-labeled tree~\((\mathcal{T},\lambda)\)
from the so-generated partition.

Let \((\mathcal{T},\lambda)\) be a leaf-labeled tree and
\(\phi=\phi_{(\lambda,\preceq)}\) be the full labeling of
\(\mathcal{T}\) produced by Algorithm~\ref{alg:label:network}.
The multiset of labels of the leaves of \((\mathcal{T},\lambda)\)
is denoted by \(M_{(\mathcal{T},\lambda)}\) and the multiset containing the 
labels of all of the vertices, except the root, of $\mathcal{T}$
is denoted by \(P_{(\mathcal{T},\lambda,\preceq)}\).
In addition, put
\[\Pi_{(\mathcal{T},\lambda,\preceq)} = \{ F_{(u,\phi)} \,:\, u \mbox{ is an interior vertex of } \mathcal{T} \},\]
implying that \(\Pi_{(\mathcal{T},\lambda,\preceq)}\) is a partition
of~\(P_{(\mathcal{T},\lambda,\preceq)}\).
As an example, consider the leaf-labeled tree~\((\mathcal{T},\lambda)\)
in Figure~\ref{fig:iso:same:label}(a) for which
we obtain the multisets
\begin{align*}
M_{(\mathcal{T},\lambda)} &= \{2,3,3,3,3,3,5,5,5,5,9\},\\
P_{(\mathcal{T},\lambda,\preceq)} &= \{1,1,1,1,2,3,3,3,3,3,4,5,5,5,5,6,7,8,9,10,11\}
\end{align*}
of labels and the partition 
\begin{align*}
\Pi_{(\mathcal{T},\lambda,\preceq)} = \{&\{2,3\},\{3,5\},\{3,5\},\{3,5\},\{3,5\},\{1,4\},\\
&\{1,1\},\{1,9\},\{6\},\{8,10\},\{7,11\}\}.
\end{align*} 

We want to reconstruct the leaf-labeled tree \((\mathcal{T},\lambda)\)
from the partition \(\Pi_{(\mathcal{T},\lambda,\preceq)}\). The following
lemma will be our starting point.

\begin{lemma}
\label{lem:number:of:leaves}
Let \((\mathcal{T},\lambda)\) be a leaf-labeled tree. Then
\[|P_{(\mathcal{T},\lambda,\preceq)}| + 1 - |\Pi_{(\mathcal{T},\lambda,\preceq)}|\]
equals the number of leaves of~\(\mathcal{T}\).
\end{lemma}

\pf
By definition, \(|P_{(\mathcal{T},\lambda,\preceq)}| + 1\) equals the
number of all vertices of~\(\mathcal{T}\) and
\(|\Pi_{(\mathcal{T},\lambda,\preceq)}|\) equals the number of
interior vertices of~\(\mathcal{T}\).
\epf

\begin{figure}
\centering
\includegraphics[scale=1.0]{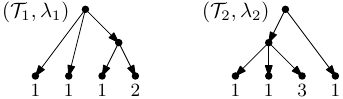}
\caption{Two non-isomorphic leaf-labeled trees~\((\mathcal{T}_1,\lambda_1)\)
and~\((\mathcal{T}_2,\lambda_2)\) with
\(\Pi_{(\mathcal{T}_1,\lambda_1,\preceq)} = \Pi_{(\mathcal{T}_2,\lambda_2,\preceq)}
= \{\{1,2\},\{1,1,3\}\}\) (independently of the choice
of \(\preceq\) in Algorithm~\ref{alg:label:network}).}
\label{fig:trees:same:partition}
\end{figure}

So, we can extract some information about \((\mathcal{T},\lambda)\)
from \(\Pi_{(\mathcal{T},\lambda,\preceq)}\), but,
as can be seen in Figure~\ref{fig:trees:same:partition},
without any further assumptions, the partition
\(\Pi_{(\mathcal{T},\lambda,\preceq)}\) does not uniquely determine the
leaf-labeled tree \((\mathcal{T},\lambda)\).

A multiset \(A \in \mathbb{M}\) is \emph{gap-less} if
\(A(x) \geq 1\) for all \(1 \leq x \leq \max(A)\).
Let~\(\preceq\) be an ordering of the multisets in~\(\mathbb{M}\).
A partition \(\Pi\) of a multiset \(P \in \mathbb{M}\)
is \(\preceq\)-\emph{tree-generated} if there exists a
leaf-labeled tree \((\mathcal{T},\lambda)\) with
\(M_{(\mathcal{T},\lambda)}\) gap-less, 
\(P = P_{(\mathcal{T},\lambda,\preceq)}\) and
\(\Pi = \Pi_{(\mathcal{T},\lambda,\preceq)}\).
Moreover, for a partition~\(\Pi\) of \(P \in \mathbb{M}\), let
\(n_{(P,\Pi)}\) denote the element of~\(P\)
at position \(|P|+1-|\Pi|\) when \(P\) is sorted
non-decreasingly. As an example, consider the leaf-labeled
tree \((\mathcal{T},\lambda)\) in Figure~\ref{fig:trees:same:partition}(a).
We have \(M_{(\mathcal{T},\lambda)}=\{1,1,1,2\}\), which is gap-less,
and \(P=P_{(\mathcal{T},\lambda,\preceq)} = \{1,1,1,2,3\}\), with elements
written in non-decreasing order. Hence, the partition
\[\Pi = \Pi_{(\mathcal{T},\lambda,\preceq)} = \{\{1,2\},\{1,1,3\}\}\]
of~\(P\) is \(\preceq\)-tree-generated and \(n_{(P,\Pi)} = 2\). 

\begin{lemma}
\label{lem:tree:generated:partition}
Let \(\Pi\) be a \(\preceq\)-tree-generated partition
of \(P \in \mathbb{M}\) and \((\mathcal{T},\lambda)\) be
a leaf-labeled tree with \(P=P_{(\mathcal{T},\lambda,\preceq)}\)
and \(\Pi = \Pi_{(\mathcal{T},\lambda,\preceq)}\). Then
\[M_{(\mathcal{T},\lambda)}(x) = \begin{cases}
P(x) &\text{if} \ 1 \leq x \leq n_{(P,\Pi)}\\
0 &\text{if} \ x > n_{(P,\Pi)}.
\end{cases}\]
\end{lemma}

\pf
By Lemma~\ref{lem:number:of:leaves}, the number of leaves
of \((\mathcal{T},\lambda)\) equals \(|P|+1-|\Pi|\).
Since \(M_{(\mathcal{T},\lambda)}\) is gap-less,
the labels assigned by Algorithm~\ref{alg:label:network}
to interior vertices of \((\mathcal{T},\lambda)\) are
strictly larger than any element of \(M_{(\mathcal{T},\lambda)}\).
Thus, extracting the first \(|P|+1-|\Pi|\) elements from
the non-decreasingly sorted multiset \(P\) yields
the elements of \(M_{(\mathcal{T},\lambda)}\), with
\(n_{(P,\Pi)}\) being the largest element of~\(M_{(\mathcal{T},\lambda)}\).
\epf

We are now ready to present Algorithm~\ref{alg:mulfromtreepartition}
which reconstructs a leaf-labeled tree from a
\(\preceq\)-tree-generated partition.

\begin{algorithm}[h]
\caption{Compute a leaf-labeled tree from a \(\preceq\)-tree-generated partition $\Pi$ of a multiset $P \in \mathbb{M}$}
\label{alg:mulfromtreepartition}
\begin{algorithmic}[1]
\Procedure{ComputeLeafLabeledTree}{$P$, $\Pi$}
\State $n \gets n_{(P,\Pi)}$ \Comment{largest label of a leaf}
\State $M \gets \{x \in P : x \leq n\}$ \Comment{multiset of leaf labels} 
\State $V \gets$ set of \(|M|\) isolated vertices bijectively labeled by elements of~$M$
\State $E \gets \emptyset$ \Comment{initialization of arc set}
\State $A \gets M$ \Comment{multiset of labels of vertices currently having in-degree~0}
\While{$\Pi \neq \emptyset$}
   \State $\Pi' \gets$ minimal elements with respect to $\preceq$ in $\{F \in \Pi  : F \subseteq A\}$
   \State $n \gets n+1$ \Comment{next value used as a label}
   \ForAll{$F \in \Pi'$}
      \State add a new vertex $u$ to $V$ and label $u$ by $n$
      \State $A(n) \gets A(n) + 1$
      \ForAll{$x \in F$}
         \State select $v \in V$ with in-degree~0 and labeled by~$x$
         \State add arc $(u,v)$ to $E$
         \State $A(x) \gets A(x) - 1$
      \EndFor 
   \EndFor
   \State $\Pi \gets \Pi - \Pi'$
\EndWhile
\State remove all labels of interior vertices \Comment{to obtain a leaf-labeled tree}
\EndProcedure
\end{algorithmic}
\end{algorithm}

\begin{theorem}
\label{thm:alg:reconstruct:tree}
Let~\(\Pi\) be a \(\preceq\)-tree-generated partition of a
multiset $P \in \mathbb{M}$ and \((\mathcal{T},\lambda)\)
a leaf-labeled tree with \(P=P_{(\mathcal{T},\lambda,\preceq)}\) and 
\(\Pi=\Pi_{(\mathcal{T},\lambda,\preceq)}\).
Then, on input \(P\) and~\(\Pi\),
Algorithm~\ref{alg:mulfromtreepartition} computes a
leaf-labeled tree that
is isomorphic to~\((\mathcal{T},\lambda)\).
\end{theorem}

\pf
By Lemma~\ref{lem:tree:generated:partition}
the set \(M\) of leaf labels of \((\mathcal{T},\lambda)\)
is uniquely determined by \(P\) and \(\Pi\).
Moreover, since \(M\) is gap-less, the set \(P\) is also
gap-less and interior vertices of \(\mathcal{T}\) are
labeled by Algorithm~\ref{alg:label:network} using the values
\(n_{(P,\Pi)}+1,\dots,\max(P)\).

We show that the following invariant holds for the while-loop
in Algorithm~\ref{alg:mulfromtreepartition}:
The current DAG \((V,E)\) can be isomorphically embedded into
\(\mathcal{T}\) by the injective map~\(f\) that assigns to each vertex~\(v\)
currently in~\(V\) a vertex of~\(\mathcal{T}\) with the same label as~\(v\). 
This invariant clearly holds before the first iteration
of the while-loop in Algorithm~\ref{alg:mulfromtreepartition}. 

Consider an iteration of the while~loop 
in Algorithm~\ref{alg:mulfromtreepartition} and assume that
the invariant holds before it. Then the multisets
\(F \in \Pi'\) selected in Line~8 of Algorithm~\ref{alg:mulfromtreepartition}
must coincide with the multisets \(F_{(w,\phi)}\)
considered in Line~7 of Algorithm~\ref{alg:label:network}.
This ensures that the construction within the for-loop
in Line~10 of Algorithm~\ref{alg:mulfromtreepartition} can be
performed and that the injective map \(f\) can be
extended to the vertices~\(u\) added in Line~11
of Algorithm~\ref{alg:mulfromtreepartition} by assigning
to them the vertices in~\(W'\) 
considered in Line~7 of Algorithm~\ref{alg:label:network}.

It follows that the invariant also holds after the
last iteration of the while~loop 
in Algorithm~\ref{alg:mulfromtreepartition}, implying
that the output of Algorithm~\ref{alg:mulfromtreepartition}
is a leaf-labeled tree that is isomorphic to \((\mathcal{T},\lambda)\).
\epf

\begin{remark}
\label{rem:decide:tree:generated}
Assuming that the outcome of
the comparison of \(A,B \in \mathbb{M}\) with respect to~\(\preceq\)
can be computed in polynomial time, Algorithm~\ref{alg:mulfromtreepartition}
runs in polynomial time. Under the same assumption,
Algorithm~\ref{alg:mulfromtreepartition} can be adapted to
decide in polynomial time (for a fixed ordering~\(\preceq\) of~\(\mathbb{M}\))
if a given partition~\(\Pi\) of \(P \in \mathbb{M}\) is
\(\preceq\)-tree-generated. 
\end{remark}

\section{Labeling-consistent orderings of $\mathbb{M}$}
\label{sec:char:tree:generated:partitions}

Ideally, we would now like to
characterize \(\preceq\)-tree-generated partitions.
However, this appears to be 
difficult without placing some restriction on~\(\preceq\).
The main obstacle is that sorting the elements in a
\(\preceq\)-tree-generated partition of some \(P \in \mathbb{M}\)
non-decreasingly by~\(\preceq\) does not
necessarily correspond to the order in which the elements of
the partition are processed by Algorithm~\ref{alg:mulfromtreepartition}
(cf. Line~8).
In~\cite{francis2023labellable}, where a similar challenge
arises, this is overcome by
selecting, for each network, a suitable fixed order.
In this section, we show that by placing a relatively mild assumption
on the ordering~\(\preceq\) (labeling-consistency)
we can characterize \(\preceq\)-tree-generated partitions for these
orderings.

\begin{figure}
\centering
\includegraphics[scale=1.0]{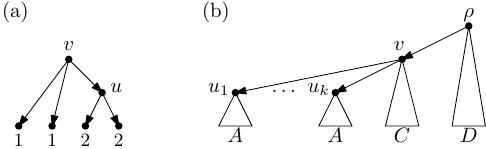}
\caption{(a) A leaf-labeled tree illustrating that the lexicographic
ordering of \(\mathbb{M}\) is not labeling-consistent.
(b) The structure of the leaf-labeled tree used in the proof
of Lemma~\ref{lem:char:labeling:consistent}. Each triangle
represents a set of leaves that are bijectively labeled
by the elements of the multiset below the triangle.}
\label{ex:not:labeling:consistent}
\end{figure}

An ordering~\(\preceq\) of~\(\mathbb{M}\) is \emph{labeling-consistent}
if, for all leaf-labeled trees \((\mathcal{T}=((V,E),\rho),\lambda)\) with
\(M_{(\mathcal{T},\lambda)}\) gap-less and all interior vertices
\(u,v \in V\),
\begin{equation}
\label{eq:labeling:consistent}
\phi(u) \leq \phi(v)
\ \text{implies} \
F_{(u,\phi)} \preceq F_{(v,\phi)},
\end{equation}
where \(\phi = \phi_{(\lambda,\preceq)}\).
Note that the lexicographic ordering~\(\preceq\) of \(\mathbb{M}\)
defined in Section~\ref{sec:full:labelings} is not labeling-consistent,
as can be seen from the leaf-labeled tree \((\mathcal{T},\lambda)\) in
Figure~\ref{ex:not:labeling:consistent}(a) for which we have
\(\phi(u) = 3 \leq 4 = \phi(v)\) but
\(F_{(u,\phi)} = \{2,2\} \not \preceq \{1,1,3\} = F_{(v,\phi)}\).
An example of an ordering of~\(\mathbb{M}\) that is 
labeling-consistent (see Corollary~\ref{cor:anti:lex:labeling:consistent}
below) is the \emph{anti-lexicographic ordering} defined by putting
\(M \preceq M'\) for \(M,M' \in \mathbb{M}\)
if either \(M=M'\) or \(M(x) < M'(x)\)
for the largest \(x \in \mathbb{N}^*\) with \(M(x) \neq M'(x)\).
This ordering is also considered in~\cite{erdHos1989applications}
(for finite non-empty subsets of~\(\mathbb{N}^*\)).
The following lemma characterizes the labeling-consistent
orderings of~\(\mathbb{M}\).

\begin{lemma}
\label{lem:char:labeling:consistent}
An ordering~\(\preceq\) of \(\mathbb{M}\) is labeling-consistent
if and only if
\begin{equation}
\label{eq:char:labeling:consistent}
\max(A) < \max(B) \ \text{implies} \ A \prec B
\end{equation}
for all \(A,B \in \mathbb{M}\).
\end{lemma}

\pf
First assume that~\(\preceq\) satisfies~\eqref{eq:char:labeling:consistent}.
We need to show that~\(\preceq\) is labeling-consistent.
Let \((\mathcal{T},\lambda)\)
be a leaf-labeled tree with \(M_{(\mathcal{T},\lambda)}\) gap-less
and \(u,v\) be interior vertices of \(\mathcal{T}\). Put
\(\phi = \phi_{(\lambda,\preceq)}\) and assume without loss of
generality \(\phi(u) \leq \phi(v)\). We consider two cases.

\textsl{Case 1}: In the iteration of the while-loop in
Algorithm~\ref{alg:label:network} where \(u\) is assigned
its label \(\phi(u)\), all children of \(v\) have been labeled
already. Then, by the construction of the set \(W'\) in 
Line~7 of Algorithm~\ref{alg:label:network}, we must have
\(F_{(u,\phi)} \preceq F_{(v,\phi)}\), as required.

\textsl{Case 2}: In the iteration of the while-loop in
Algorithm~\ref{alg:label:network} where \(u\) is assigned
its label \(\phi(u)\), there exists a child \(w\) of \(v\)
that has not been assigned a label yet. Since
\(M_{(\mathcal{T},\lambda)}\) is gap-less, we have
\(\max(F_{(u,\phi)}) < \phi(u) \leq \phi(w) \leq \max(F_{(v,\phi)})\).
Thus, by the assumption that~\(\preceq\)
satisfies~\eqref{eq:char:labeling:consistent},
\(F_{(u,\phi)} \prec F_{(v,\phi)}\), as required.

Next assume that there exist \(A,B \in \mathbb{M}\)
with \(\max(A) < \max(B)\) and \(B \prec A\).
We need to show that~\(\preceq\) is not labeling-consistent.
Put \(m = \max(B)\) and \(k = B(m)\).
Consider the multisets \(C,D \in \mathbb{M}\) with
\[
C(x) = \begin{cases}
B(x) &\text{if} \ x < m\\
0 &\text{if} \ x \geq m
\end{cases} \quad \text{and} \quad
D(x) = \begin{cases}
1 &\text{if} \ x < m\\
0 &\text{if} \ x \geq m
\end{cases}
\]
We consider the leaf-labeled tree \((\mathcal{T},\lambda)\)
whose structure is depicted in Figure~\ref{ex:not:labeling:consistent}(b).
\(\mathcal{T}\) has \(k \cdot |A| + |C| + |D|\) leaves.
More specifically, each of the interior vertices
\(u_i\), \(1 \leq i \leq k\), has \(|A|\) children that are
all leaves and bijectively labeled by the elements of~\(A\).
The interior vertex \(v\) has \(k + |C|\) children, among them
\(|C|\) leaves that are bijectively labeled by the elements of~\(C\).
Finally, the root of \(\mathcal{T}\) has \(1 + |D|\) children,
among them
\(|D|\) leaves that are bijectively labeled by the elements of~\(D\).
Then, by the definition of~\(D\),
\(M_{(\mathcal{T},\lambda)}\) is gap-less.

Put \(\phi = \phi_{(\lambda,\preceq)}\). Then we have
\(\phi(u_i) = m\) for \(1 \leq i \leq k\), \(\phi(v)=m+1\)
and \(\phi(\rho) = m+2\). Thus, putting \(u=u_1\),
we have \(\phi(u) < \phi(v)\) but
\(A = F_{(u,\phi)} \not \prec F_{(v,\phi)} = B\),
implying that \(\preceq\) is not labeling-consistent,
as required.
\epf

\begin{corollary}
\label{cor:anti:lex:labeling:consistent}
The anti-lexicographic ordering of \(\mathbb{M}\) is
labeling-consistent.
\end{corollary}

\pf
It follows immediately from the definition of the
anti-lexicographic ordering that~\eqref{eq:char:labeling:consistent}
holds.
\epf

For a partition~\(\Pi\) of \(P \in \mathbb{M}\) and \(F \in \Pi\),
the multiplicity of \(F\) in \(\Pi\) is denoted by \(\Pi(F)\).
Moreover, \(\underline{\Pi}\) denotes the set containing
a single copy of each \(F \in \Pi\).

\begin{theorem}
\label{char:exp:partitions}
Let \(\preceq\) be a labeling-consistent ordering of~\(\mathbb{M}\),
\(\Pi\) be a partition of \(P \in \mathbb{M}\) and
\(F_1,\dots,F_k\) be the elements of \(\underline{\Pi}\)
sorted increasingly by~\(\preceq\).
Then \(\Pi\) is \(\preceq\)-tree-generated if and only if
\begin{itemize}
\item[(i)]
\(P\) is gap-less,
\item[(ii)]
\(|\underline{\Pi}| = \max(P) - n_{(P,\Pi)} = k\),
\item[(iii)]
\(P(i+n_{(P,\Pi)}) = \Pi(F_i)\) for all \(1 \leq i \leq k\), and
\item[(iv)]
\(\max(F_i) \leq n_{(P,\Pi)} + i -1\) for all \(1 \leq i \leq k\).
\end{itemize}
\end{theorem}

\pf
First assume that \(\Pi\) is \(\preceq\)-tree-generated.
Let \((\mathcal{T},\lambda)\) be a leaf-labeled tree
with \(P = P_{(\mathcal{T},\lambda,\preceq)}\)
and \(\Pi = \Pi_{(\mathcal{T},\lambda,\preceq)}\).
Since \(\Pi\) is \(\preceq\)-tree-generated,
\(M_{(\mathcal{T},\lambda)}\) is gap-less, implying~(i).
(ii) follows from Lemma~\ref{lem:tree:generated:partition}.
(iii) follows from the definition of
\(\Pi_{(\mathcal{T},\lambda,\preceq)}\) and the assumption that~\(\preceq\)
as labeling-consistent. (iv) follows from the fact that,
since \(M_{(\mathcal{T},\lambda)}\) is gap-less,
before the \(i\)-th iteration of the while-loop in
Algorithm~\ref{alg:label:network} the maximum label
used so far is \(n_{(P,\Pi)} + i -1\).

Next assume that~\(\Pi\) satisfies Properties~(i)-(iv).
We show that Algorithm~\ref{alg:mulfromtreepartition}
on input \(P\) and \(\Pi\) produces a leaf-labeled tree
\((\mathcal{T},\lambda)\) with \(P = P_{(\mathcal{T},\lambda,\preceq)}\)
and \(\Pi = \Pi_{(\mathcal{T},\lambda,\preceq)}\).
Properties (i)-(iii) ensure that the set~\(M\)
constructed in Line~3 of Algorithm~\ref{alg:mulfromtreepartition}
can be used as the multiset of leaf-labels of \((\mathcal{T},\lambda)\).

To finish the proof, it suffices to show
that the following invariant holds for the while-loop
in Algorithm~\ref{alg:mulfromtreepartition}:
The multiset \(\{F \in \Pi: F \subseteq A\}\) is not empty. 
This invariant clearly holds before the first iteration
of the while-loop in Algorithm~\ref{alg:mulfromtreepartition}
since, by (iv), \(\max(F_1) \leq n_{(P,\Pi)} = \max(M)\).

Consider the \(i\)-th iteration, \(1 \leq i \leq k\),  of the while~loop 
in Algorithm~\ref{alg:mulfromtreepartition} and assume that
the invariant holds before it. Then \(\Pi(F_i)\) new vertices
are added in the for-loop in Line~10
of Algorithm~\ref{alg:mulfromtreepartition}.
All these new vertices are assigned~\(n_{(P,\Pi)} + i\) as their
label. Thus, by~(iv) and since \(\Pi\) is a partition of \(P\),
we must have \(F_{i+1} \subseteq A\) and, therefore,
the invariant holds also after the \(i\)-th iteration
of the while~loop in Algorithm~\ref{alg:mulfromtreepartition}.
\epf

\section{Folding and unfolding}
\label{sec:folding:unfolding}

We now shift our focus from leaf-labeled trees to 
more general leaf-labeled networks.
We call a leaf-labeled network~\((\mathcal{N},\lambda)\)
\emph{labelable} if the full labeling \(\phi = \phi_{(\lambda,\preceq)}\) of the network
produced by Algorithm~\ref{alg:label:network} is injective.
Note that in view of Theorem~\ref{thm:labeling:injective},
the choice of the ordering~\(\preceq\) has no impact
on whether or not a leaf-labeled network is labelable.
In this section, we present a characterization
of labelable networks (Theorem~\ref{thm:stable:networks}),
which we shall use in the 
next section to give a new characterization of labelable
phylogenetic networks.

\begin{figure}
\centering
\includegraphics[scale=1.0]{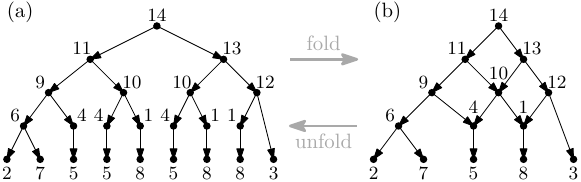}
\caption{Illustration of folding and unfolding.
(a) A fully labeled tree \((\mathcal{T},\phi)\). The
full labeling \(\phi\) is obtained by applying
Algorithm~\ref{alg:label:network} using as \(\preceq\)
the lexicographic ordering.
(b) The fully labeled network
\((\mathcal{N}',\phi')=\mathfrak{F}(\mathcal{T},\phi)\)
obtained by folding \((\mathcal{T},\phi)\). 
Unfolding \((\mathcal{N}',\phi')\) reproduces \((\mathcal{T},\phi)\).}
\label{fig:ex:fold:unfold}
\end{figure}

Let \((\mathcal{T}=((V,E),\rho),\lambda)\) be
a leaf-labeled
tree and \(\phi = \phi_{(\lambda,\preceq)}\) be the full labeling
of \(\mathcal{T}\) produced by Algorithm~\ref{alg:label:network}. 
Then \(\mathfrak{F}(\mathcal{T},\phi)\) denotes the
fully labeled network \((((V',E'),\rho'),\phi')\) obtained
as follows:
\begin{itemize}
\item
\(V' = \phi(V)\).
\item
\(E' = \{(\phi(u),\phi(v)) : (u,v) \in E\}\).
\item
\(\rho' = \phi(\rho)\).
\item
\(\phi' : V' \rightarrow \mathbb{N}^*\) with \(\phi'(k) = k\).
\end{itemize}
The fully labeled network \(\mathfrak{F}(\mathcal{T},\phi)\)
is called the \emph{folding} of \((\mathcal{T},\phi)\)
(see Figure~\ref{fig:ex:fold:unfold} for an example).
Note that in~\cite{huber2016folding} the folding of
a leaf-labeled tree \((\mathcal{T}=((V,E),\rho),\lambda)\) 
is described in terms of isomorphisms
between the leaf-labeled trees \((\mathcal{T}_u,\lambda_u)\),
\(u \in V\). In view of Lemma~\ref{lem:label:tree:isomorphic},
this is equivalent to the definition above.

Next we describe the reverse process.
Let \((\mathcal{N}=((V,E),\rho),\phi)\) be a fully labeled
network. Then \(\mathfrak{U}(\mathcal{N},\phi)\) denotes the
fully labeled tree \((((V',E'),\rho'),\phi')\) obtained
as follows:
\begin{itemize}
\item
\(V'\) consists of the set of all directed paths in
\(\mathcal{N}\) from \(\rho\) to some \(v \in V\).
\item
\((\pi,\pi')\) is an arc in \(E'\) if the directed path \(\pi'\)
arises from the directed path \(\pi\) by extending \(\pi\)
along an arc in~\(E\).
\item
\(\rho'\) is the directed path in \(\mathcal{N}\) consisting
of the single vertex \(\rho\).
\item
\(\phi' : V' \rightarrow \mathbb{N}^*\) with \(\phi'(\pi) = \phi(u)\),
where \(u\) is the last vertex in the directed path~\(\pi\). 
\end{itemize}
The fully labeled tree \(\mathfrak{U}(\mathcal{N},\phi)\)
is called the \emph{unfolding} of \((\mathcal{N},\phi)\)
(see again Figure~\ref{fig:ex:fold:unfold} for an example).

The next lemma states more precisely the idea that the process
of unfolding a fully-labeled network and the application of
Algorithm~\ref{alg:label:network} are compatible.

\begin{lemma}
\label{lem:label:reconstruction}
Let \((\mathcal{N}=((V,E),\rho),\lambda)\) be a leaf-labeled
network, \(\phi = \phi_{(\lambda,\preceq)}\) be the full labeling
of \(\mathcal{N}\) produced by Algorithm~\ref{alg:label:network},
\((\mathcal{T}',\phi')= \mathfrak{U}(\mathcal{N},\phi)\)
and \(\lambda' : L(\mathcal{T}') \rightarrow \mathbb{N}^*\) be
the leaf labeling of \(\mathcal{T}'\) obtained by putting
\(\lambda'(v') = \phi'(v')\) for all \(v' \in L(\mathcal{T}')\).
Then the full labeling \(\phi''=\phi_{(\lambda',\preceq)}\) of
\(\mathcal{T}'\) produced by Algorithm~\ref{alg:label:network}
coincides with~\(\phi'\).
\end{lemma}

\pf
The proof is by induction over the iterations of the
while loop in Algorithm~\ref{alg:label:network}.
The base case involves the leaves of \(\mathcal{T}'\)
and clearly holds.

If an interior vertex \(u'\) of \(\mathcal{T}'\) is labeled
by \(\phi''\) in an iteration of the while
loop, all children of \(u'\) have been labeled
by \(\phi''\) in some earlier iteration.
Hence, by induction, we have \(\phi''(w') = \phi'(w')\)
for all children~\(w'\) of~\(u'\).
By construction of \((\mathcal{T}',\phi')\), \(u'\)
corresponds to a directed path \(\pi\) from \(\rho\) to some
vertex \(u\) in \(\mathcal{N}\). Moreover, the
children of \(u'\) in \(\mathcal{T}'\) correspond
to the extensions of \(\pi\) by an arc from \(u\)
to a child of \(u\) in \(\mathcal{N}\). Hence,
the multiset of labels of the children of \(u'\)
in \(\mathcal{T}'\) coincides with the multiset
of labels of the children of \(u\)
in \(\mathcal{N}\). But this implies that
\(\phi(u) = \phi'(u') = \phi''(u')\), as required.
\epf

\begin{remark}
\label{rem:labeling:tree:network:consistent}
From the proof of Lemma~\ref{lem:label:reconstruction}
we see that Algorithm~\ref{alg:label:network},
when applied to \((\mathcal{N},\lambda)\) and
to \((\mathcal{T}',\lambda')\), respectively, labels,
in each iteration, vertices whose children have the same multiset of
labels.
\end{remark}

A leaf-labeled network \((\mathcal{N},\lambda)\) is called
\emph{stable} if the fully labeled network
\((\mathcal{N},\phi_{(\lambda,\preceq)})\) is isomorphic
to \(\mathfrak{F}(\mathfrak{U}(\mathcal{N},\phi_{(\lambda,\preceq)}))\) \cite{huber2016folding}.
As can be seen in Figure~\ref{fig:ex:fold:unfold}, stable networks
exist. In contrast, the leaf-labeled network in
Figure~\ref{fig:ex:leaf:labeled:network}(a)
is not stable, since unfolding it yields the
fully labeled tree in Figure~\ref{fig:iso:same:label}(b)
but folding this tree does not reproduce
the fully labeled network in Figure~\ref{fig:ex:leaf:labeled:network}(b).

We now present the aforementioned characterization of
labelable networks.

\begin{theorem}
\label{thm:stable:networks}
A leaf-labeled network \((\mathcal{N},\lambda)\) is  labelable
if and only if it is stable.
\end{theorem}

\pf
Assume that \((\mathcal{N},\lambda)\) is stable.
Then the full labeling \(\phi_{(\lambda,\preceq)}\) 
of \(\mathcal{N}\) produced by Algorithm~\ref{alg:label:network}
must be injective since the full labeling
of \(\mathfrak{F}(\mathfrak{U}(\mathcal{N},\phi_{(\lambda,\preceq)}))\)
is injective by the definition of the folding.
Hence, \((\mathcal{N},\lambda)\) is labelable.

Next assume that \((\mathcal{N},\lambda)\) is labelable.
Then the full labeling \(\phi = \phi_{(\lambda,\preceq)}\) 
of \(\mathcal{N}\) produced by Algorithm~\ref{alg:label:network}
is injective by definition. Then the required isomorphism
\(f\) between \((\mathcal{N},\lambda)\) and 
\(\mathfrak{F}(\mathfrak{U}(\mathcal{N},\phi_{(\lambda,\preceq)}))\)
is obtained by putting \(f(u) = \phi(u)\) for all
vertices \(u\) of \(\mathcal{N}\).
\epf

\section{Classes of phylogenetic networks through an expanding partition lens}
\label{sec:network:classes}

A leaf-labeled network \((\mathcal{N},\lambda)\) is a
\emph{phylogenetic network} 
if all leaves of~\(\mathcal{N}\) have in-degree~1 and
\(\lambda\) is a bijection between \(L(\mathcal{N})\)
and \([n]\) for \(n=|L(\mathcal{N})|\)~\cite{francis2023labellable}. 
We also call such a leaf-labeled network a
\emph{phylogenetic network on}~\([n]\) for short.
In \cite{francis2023labellable} a bijection 
is established between the set of labelable 
phylogenetic networks on~\([n]\) and expanding covers
(defined below). In this section, via foldings, 
we extend this bijective correspondence to include stable 
phylogenetic networks and a certain class of partitions of
multisets.

We first recall a key definition from \cite{francis2023labellable}.
An \emph{expanding cover}
is a collection \(\mathcal{C}\) of non-empty subsets
of~\([m]\) for some \(m \in \mathbb{N}^*\) such that
\begin{itemize}
\item[(EC1)]
\(\bigcup_{C \in \mathcal{C}} C = [m]\),
\item[(EC2)]
\(n_{\mathcal{C}} = m - |\mathcal{C}| + 1 \geq 1\),
\item[(EC3)]
For all \(x \in [n_{\mathcal{C}}]\) there exists a unique \(C \in \mathcal{C}\)
with \(x \in C\), and
\item[(EC4)]
For all \(1 \leq i \leq |\mathcal{C}|\) there exist at least
\(i\) sets \(C \in \mathcal{C}\) with \(C \subseteq [n_{\mathcal{C}}+i-1]\).
\end{itemize}

Now, let~\(\preceq\) be an ordering of \(\mathbb{M}\).
A \(\preceq\)-\emph{expanding partition}
is a \(\preceq\)-tree-generated partition~\(\Pi\) of
some \(P \in \mathbb{M}^*\) such that
\begin{itemize}
\item[(EP1)]
for all \(F \in \Pi\) and all \(x \in F\) we have \(F(x)=1\), and
\item[(EP2)]
For all \(F,F' \in \Pi\) with \(F \neq F'\) we have 
\(F \cap F' \cap [n_{(P,\Pi)}] = \emptyset\).
\end{itemize}
Since~\(P\) is determined by~\(\Pi\), we put
\(n_{\Pi} = n_{(P,\Pi)}\) for such a partition.
We now extend the correspondence given
in \cite[Theorem 4.4]{francis2023labellable}.

\begin{theorem}
\label{thm:bijections:labelable}
Let~\(n \in \mathbb{N}^*\) and~\(\preceq\) an
ordering of~\(\mathbb{M}\).
The following sets are in bijective correspondence:
\begin{itemize}
\item[(i)] The set of labelable phylogenetic networks on~\([n]\).
\item[(ii)] The set of stable phylogenetic networks on~\([n]\).
\item[(iii)] The set of expanding covers~\(\mathcal{C}\) with
\(n_{\mathcal{C}} = n\).
\item[(iv)] The set of \(\preceq\)-expanding partitions~\(\Pi\) with
\(n_{\Pi} = n\).
\end{itemize}
\end{theorem}

\pf
The bijective correspondence between the sets in~(i) and~(ii)
follows from Theorem~\ref{thm:stable:networks}.

The bijective correspondence between the sets in~(i) and~(iii)
is established in \cite[Thm.~4.4]{francis2023labellable}.

To establish the bijective correspondence between the sets in~(ii) and~(iv),
we consider the map~\(p\) that associates to each stable phylogenetic
network \((\mathcal{N}',\lambda')\) on \([n]\) the
partition~\(\Pi_{(\mathcal{T},\lambda,\preceq)}\) of~\(P_{(\mathcal{T},\lambda,\preceq)}\),
where \((\mathcal{T},\lambda)\) is the leaf-labeled tree
obtained by unfolding the fully labeled network 
\((\mathcal{N}',\phi_{(\lambda',\preceq)})\) and then restricting
the labeling of the fully labeled tree
\(\mathfrak{U}(\mathcal{N}',\phi_{(\lambda',\preceq)})\)
to its leaves. By Lemma~\ref{lem:label:reconstruction},
the fully labeled tree~\((\mathcal{T},\phi_{(\lambda,\preceq)})\)
coincides with \(\mathfrak{U}(\mathcal{N}',\phi_{(\lambda',\preceq)})\).
Moreover, since \((\mathcal{N}',\lambda')\) is stable and,
therefore, labelable, the full labeling
\(\phi_{(\lambda',\preceq)}\) of \(\mathcal{N}'\) is, by definition,
injective. But this implies, by the definition of
\(\mathfrak{U}(\mathcal{N}',\phi_{(\lambda',\preceq)})\),
that~\(\Pi_{(\mathcal{T},\lambda,\preceq)}\) satisfies~(EP1).
In addition, since \((\mathcal{N}',\lambda')\) is a phylogenetic
network on~\([n]\), \(n_{\Pi_{(\mathcal{T},\lambda,\preceq)}} = n\) and
\(\Pi_{(\mathcal{T},\lambda,\preceq)}\) satisfies~(EP2).
Thus, \(p\)~maps every stable phylogenetic network
on~\([n]\) to a \(\preceq\)-expanding partition~\(\Pi\)
with \(n_{\Pi} = n\).

It remains to show that the map~\(p\) is bijective.
It follows from the definition of stability of a leaf-labeled network
together with Theorem~\ref{thm:alg:reconstruct:tree}
that~\(p\) is injective. To show that~\(p\) is also surjective,
consider a \(\preceq\)-expanding partition~\(\Pi\)
of \(P \in \mathbb{M}\) with \(n_{\Pi} = n\).
Let \((\mathcal{T},\lambda)\) be the leaf-labeled
tree produced by Algorithm~\ref{alg:mulfromtreepartition}
for~\(\Pi\) and~\(P\). In addition, let
\((\mathcal{N}',\lambda')\) be the leaf-labeled network
obtained by folding the fully labeled tree
\((\mathcal{T},\phi_{(\lambda,\preceq)})\) and then
restricting the labeling of the fully labeled
network \(\mathfrak{F}(\mathcal{T},\phi_{(\lambda,\preceq)})\)
to its leaves. Since \(n_{\Pi} = n\), \(\lambda'\) is
a bijection between \(n\) and the set of leaves of \(\mathcal{N}'\).
Moreover, since~\(\Pi\) satisfies~(EP2), all leaves 
of \(\mathcal{N}'\) have in-degree~1. Hence,
\((\mathcal{N}',\lambda')\) is a phylogenetic network on~\([n]\).

We claim that
\(\mathfrak{U}(\mathcal{N}',\phi_{(\lambda',\preceq)})\) is isomorphic
to \((\mathcal{T},\phi_{(\lambda,\preceq)})\). 
Note that this claim
immediately implies that \((\mathcal{N}',\lambda')\) is stable
and \(p(\mathcal{N}',\lambda') = \Pi\), as required.
Put \(\phi = \phi_{(\lambda,\preceq)}\) and let~\(\rho\) denote
the root of~\(\mathcal{T}\).
To establish the claim, it suffices to show,
in view of the definition of the unfolding of a leaf-labeled network,
that, for any two directed paths 
\(\rho = v_1,\dots,v_k\) and
\(\rho = u_1,\dots,u_\ell\)
in \((\mathcal{T},\phi_{(\lambda,\preceq)})\)
with \(v_k \neq u_\ell\),
the sequences
\(\phi(v_1),\dots,\phi(v_k)\) and
\(\phi(u_1),\dots,\phi(u_\ell)\) of vertex labels 
differ. Assume for a contradiction that there exist
two such paths with the same sequence of vertex labels.
Then we have \(k=\ell\) and,
since \(v_1 = \rho = u_1\) and \(v_k \neq u_k\),
there exists a unique \(1 \leq i < k\) with
\(v_i = u_i\) and \(v_{i+1} \neq u_{i+1}\). 
By our assumption we have \(\phi(v_{i+1}) = \phi(u_{i+1})\),
implying that the multiplicity of the label \(\phi(v_{i+1})\)
in \(F_{(v_i,\phi)}\) is at least~2. This is a contradiction
to the fact that, by construction, \(F_{(v_i,\phi)} \in \Pi\)
and \(\Pi\) satisfies~(EP1).
\epf

\begin{remark}
\label{rem:cover:vs:partition}
The expanding cover associated with a phylogenetic network
arises in a similar way as a \(\preceq\)-tree-generated partition
from a leaf-labeled tree (see \cite{francis2023labellable} for
details). In particular,
it follows from the proof of Theorem~\ref{thm:bijections:labelable}
that the expanding cover~\(\mathcal{C}\) that corresponds
to an expanding partition~\(\Pi\) coincides with~\(\underline{\Pi}\).
\end{remark}

In \cite{francis2024phylogenetic,francis2023labellable}
some classes of phylogenetic networks are characterized
in terms of their associated expanding covers.
In view of Remark~\ref{rem:cover:vs:partition}, these can
be translated quite easily into characterizations in terms
of their associated expanding partitions. 
We conclude this section by illustrating this
process for two classes of phylogenetic networks.

A phylogenetic network is \emph{non-degenerate}
(again as defined in \cite{francis2023labellable})
if it has no vertex with both in-degree and out-degree equal to~1
and also no vertex with both in-degree and out-degree strictly
larger than~1.

\begin{corollary}
\label{cor:char:non:degenerate:networks}
Let~\(n \in \mathbb{N}^*\) and~\(\preceq\) a
labeling-consistent ordering of~\(\mathbb{M}\). The following sets
are in bijective correspondence:
\begin{itemize}
\item[(i)]
The set of non-degenerate labelable phylogenetic networks on~\([n]\).
\item[(ii)]
The set of \(\preceq\)-expanding partitions~\(\Pi\) with
\(n_{\Pi} = n\) satisfying, for all \(1 \leq i < k = |\underline{\Pi}|\),
that
\begin{itemize}
\item
\(|F_i| > 1\) implies that are are no \(F,F' \in \Pi\) with
\(F \neq F'\), \(F(n+i) \geq 1\) and \(F'(n+i) \geq 1\), and 
\item
\(|F_i| = 1\) implies that there are \(F,F' \in \Pi\) with
\(F \neq F'\), \(F(n+i) \geq 1\) and \(F'(n+i) \geq 1\),
\end{itemize}
where \(F_1,\dots,F_k\) are the elements of \(\underline{\Pi}\)
sorted increasingly by~\(\preceq\).
\end{itemize}
\end{corollary}

\pf
This is a translation of \cite[Thm.~5.1]{francis2023labellable}.
\epf

A phylogenetic network is a \emph{tree-child network}
if every interior vertex has a child with in-degree~1.
It is shown in \cite[Thm.~6.4]{francis2023labellable}
that tree-child networks are labelable.

\begin{corollary}
\label{cor:char:tree:child:networks}
Let~\(n \in \mathbb{N}^*\) and~\(\preceq\) an
ordering of~\(\mathbb{M}\). The following sets
are in bijective correspondence:
\begin{itemize}
\item[(i)]
The set of tree-child networks on~\([n]\).
\item[(ii)]
The set of \(\preceq\)-expanding partitions~\(\Pi\) with
\(n_{\Pi} = n\) and the property that,
for all \(F \in \Pi\), there exists \(x \in F\)
such that \(F'(x) = 0\) for all \(F' \in \Pi\) with \(F' \neq F\).
\end{itemize}
\end{corollary}

\pf
This is a translation of \cite[Thm.~4.1]{francis2024phylogenetic}.
\epf

\section{Concluding remarks}
\label{sec:discuss}

We conclude by suggesting some possible future directions.
In the last section we characterized when 
\(\preceq\)-expanding partitions
correspond to tree-child networks. It could be interesting
to find similar characterizations for other 
classes of networks such as tree-based networks,
and tree-sibling networks (see e.g. \cite{kong2022classes} for 
definitions of these and other classes of networks).
Note that this question has been studied in the context of 
expanding covers in \cite{francis2024phylogenetic}
(see e.g. Theorems 3.2 and 6.2 for tree-based and tree-sibling networks, respectively).
More generally, it could also be interesting to study 
network classes that are not labelable from the labeling perspective. 
In particular, 
any DAG can be topologically sorted, 
and so we could ask under what assumptions is it possible to
recover \emph{any} leaf-labeled network
from data produced by a sorting/labeling of its vertices?

In a different direction, in \cite{francis2022brauer} tree labelings 
were used to produce a semi-group structure on the set of binary phylogenetic 
trees via the multiplication of Brauer diagrams. In that paper, it 
is also asked if it is possible to represent phylogenetic networks 
within a diagram semigroup framework (see p.~23). It would be interesting to investigate
if our results could be used to help answer this question. Moreover,
in a similar vein, in \cite{erdHos1993new,francis2022brauer}
bijections are given for rooted semi-labeled forests, and it 
might be worthwhile to see if our results can be developed in the multi-labeled setting.
This could be interesting since it might help shed light 
on counting multi-labeled trees/forests where relatively little is known 
(see \cite{czabarka2013generating} for some example results).

Finally, encoding phylogenetic
trees using labelings has recently become 
highly topical in phylogenetics since such labelings can 
be used to encode trees in terms of vectors. These so-called \emph{vector encodings} 
are highly amenable to processing with machine learning algorithms
which can be useful when dealing with massive collections of trees (see e.g. \cite{chauve2025vector,penn2025phylo2vec,richman2025vector}).
It could be useful to adapt the approach we detail here to 
derive similar encodings for stable phylogenetic networks to, for example,
help with their fast comparison or in the computation of 
consensus networks. It would also be interesting to understand how 
comparison of vector encodings for (subclasses of) stable 
networks would compare with 
other phylogenetic network metrics (see e.g. \cite{cardona2008comparison}), a topic which has recently been studied for phylogenetic trees in \cite{linz2025order}. \\

\noindent{\bf Acknowledgment:}
VM would like to thank the Institute for Computational and Experimental Research in Mathematics in Providence, RI, where -- during the semester program on “Theory, Methods, and Applications of Quantitative Phylogenomics” -- he did some of the research presented in this paper. The authors also thank Katharina Huber for pointing out the connection mentioned above between labelings and vector encodings of phylogenetic trees.

\bibliographystyle{plain}
\bibliography{matchings}

\end{document}